**Direct fabrication and IV characterization of buried graphitic channels in diamond with MeV ion implantation**

P. Olivero[1], G. Amato[2], F. Bellotti[2], S. Borini[2], A. Lo Giudice[1], F. Picollo[1], E. Vittone[1]*

[1] Experimental Physics Department / Centre of Excellence “Nanostructured Interfaces and Surfaces”, University of Torino, and INFN sez. Torino Via P. Giuria 1, 10125 Torino, Italy

[2] Quantum Research Laboratory, Istituto Nazionale di Ricerca Metrologica, Strada delle Cacce 91, 10135 Torino, Italy

* corresponding author (ettore.vittone@unito.it)

**Abstract**

In the present work we report about the investigation of the conduction mechanism of $sp_2$ carbon micro-channels buried in single crystal diamond. The structures are fabricated with a novel technique which employs a MeV focused ion-beam to damage diamond in conjunction with variable thickness masks. This process changes significantly the structural proprieties of the target material, because the ion nuclear energy loss induces carbon conversion from $sp_3$ to $sp_2$ state mainly at the end of range of the ions (few micrometers). Furthermore, placing a mask with increasing thickness on the sample it is possible to modulate the channels depth at their endpoints, allowing their electrical connection with the surface. A single-crystal HPHT diamond sample was implanted with 1.8 MeV $He^+$ ions at room temperature, the implantation fluence was set in the range $2.1\cdot10^{16}$ - $6.3\cdot10^{17}$ ions $cm^{-2}$, determining the formation of buried micro-channels at ~3 µm. After deposition of metallic contacts at the channels’ endpoints, the electrical characterization was performed measuring the I-V curves at variable temperatures in the 80-690 K range. The Variable Range Hopping model was used to fit the experimental data in the ohmic regime, allowing the estimation of characteristic parameters such as the density of localized states at the Fermi level. A value of $5.5\cdot10^{17}$ states $cm^{-3}$ $eV^{-1}$ was obtained, in satisfactory agreement with values previously reported in literature. The power-law dependence between current and voltage is consistent with the space charge limited mechanism at moderate electric fields.

## Introduction

In the 70s, the pioneering work of Vavilov et al. [1] triggered a series of studies on the effect of ion induced damage on the electrical conduction properties of diamond. Since the early years, Hauser et al. demonstrated that the electrical properties of ion-implanted diamond layers were similar to those of amorphous carbon produced by sputtering graphite [2, 3]. In the following decades, extensive work has been conducted on this topic, elucidating the characteristic conduction mechanisms in structurally damaged diamond. Such processes can be divided into two characteristic regimes: when the damage density exceeds a critical level (usually referred as "graphitization threshold" [4]), after high temperature thermal annealing the damaged structure exhibits ohmic conductivity. Examples of such behavior in buried conductive channels created with MeV ion deep implantations can be found in [5] and [6]. On the other hand, when the damage density does not reach a critical value and/or the sample is not annealed at high temperature, the conduction mechanism in the resulting amorphized layers can be described with more articulated models based on variable range hopping, as discovered by Hauser et al. [2, 3] and investigated in further details by Prins [7, 8]. While in the former case the nature of the conductivity is unambiguously attributed to metallic charge transport in a graphite-like medium, in the latter case more articulate models must be adopted, as reported in detailed works by Kalish et al. [9, 10, 11] and other groups [12, 13, 14], where a number of characteristic energies for hopping transitions were extracted from the temperature dependence of the resistivity. While the adoption of different models to the description of temperature-dependent current-voltage characteristics of amorphized diamond is still under investigation, it is worth stressing that a very limited amount of works were focused on the fabrication of buried conductive channels in diamond by means of high energy MeV ions. Examples of such works can be found in [5, 6] for graphitization at high damage levels, and in [15] for sub-graphitization damage levels.

In the present work we report about the direct fabrication of buried damaged microchannels in single crystal diamond with MeV ion implantation through variable-thickness masks, allowing the formation of the layers at variable depths and thus their electrical connection with the sample surface at their endpoints. The channels were characterized with current-voltage measurements, both at room temperature and at variable temperature in the 80 - 690 K range.

## Experimental

The process of damage induced by energetic ions in matter occurs mainly at the end of ion range, where the cross section for nuclear collisions is strongly enhanced, after the ion energy is progressively reduced by electronic collisions occurring in the initial stages of the ion path [16]. In most of the research works conducted so far, keV ion implantation was employed, resulting in the amorphization/graphitization of a layer extending from the end of range of ions to the surface of the sample [3, 8, 9, 10, 12, 13], while the formation of deep conducting layers by MeV ion implantation has been explored in a limited number of works [5, 6, 15]. The different damage profiles of $\sim 10^1$-$10^2$ keV ions with respect to $\sim 10^0$ MeV ions is shown in the SRIM [17] Monte Carlo simulation in Fig. 1 for the implantation conditions reported in [3] (70 keV C ions) and in the present work (1.8 MeV $He^+$ ions). The curves were calculated by setting a value of 50 eV for the atom displacement energy in the diamond lattice [18].

The basic concept is shown schematically in Fig. 2. The ion implantation is performed by scanning an ion beam along a linear path. As the beam scan progresses, incident ions cross a mask with increasing thickness, thus progressively reducing their range in the diamond layer. As an example, Fig. 3 shows the vacancy profile generated by 1.8 MeV $He^+$ ions in diamond after crossing different tungsten thicknesses. A tungsten thickness of the order or 3 μm is sufficient to shift the end-of-range damage peak at the diamond surface. In the present work, tungsten wires with diameter of ~300 μm were successfully employed as variable-thickness masks. Therefore, after the removal of the mask, the damaged layer is connected with the sample surface in the specific point where the mask had a thickness equal to the ion range in the masking material.

The sample under exam is a synthetic single crystal diamond produced by Sumitomo with high pressure high temperature (HPHT) method. The crystal is 3×3×1.5 $mm^3$ in size and it is classified as type Ib, having a substitutional nitrogen concentration between 10 and 100 ppm, as indicated by the manufacturer. The sample is cut along the 100 crystal direction and it is optically polished on the two opposite large faces. The sample was implanted at room temperature with a 1.8 MeV $He^+$ ion microbeam at the ion microscopy line of the AN2000 accelerator facility of the INFN Legnaro National Laboratories (INFN-LNL). Helium ions were chosen so as to exclude doping effects on the observed conductivity. The ion current was below 300 pA in order to minimize beam heating effects. The ion beam was focused to a micrometer-size spot and raster scanned to achieve a uniform fluence delivery across the implantation area.

The Rutherford Backscattering (RB) signal from the metal wires was employed both to locate the implantation area with good spatial accuracy, and to monitor the implantation fluence in real time. Seven channels (150 $\mu$m long and 40 $\mu$m wide) were implanted at increasing fluences, ranging from $2.1 \cdot 10^{16}$ cm$^{-2}$ to $6.3 \cdot 10^{17}$ cm$^{-2}$. After ion implantation, the emerging channels were contacted with Ag layers evaporated through a suitable mask.

The current-voltage (I-V) measurements were carried out at room temperature using an analytical micro-probe station (Alessi REL-4500) connected to a Semiconductor Parameter Analyzer (4145B Hewlett Packard). The lowest limit for the current detection (~0.1 nA) was determined by the conductivity of the surface due to impurities as evaluated by measuring the current flowing between two Ag electrodes not connected by damaged channels. The I-V measurements were collected by connecting the probe tips with the metallic contacts evaporated on the channels end-points.

The I-V curves at various T were collected by a Keithley 6430 source-meter, in a Janis ST-100 continuous flow cryostat equipped with a heater and a Lakeshore 331 controller, in vacuum (pressure of $1 \times 10^{-5}$ mbar).

**Results**

The first result of this study concerns the effectiveness of the method based on ion irradiation to fabricate conductive buried channels emerging at the diamond surface with variable-thickness masks. This is ascertained in Fig. 4a, which shows the I-V characteristics at room temperature relevant to the 7 channels damaged at increasing fluences. The currents measured between electrodes connected by damaged channels are significantly higher than the current flowing between two not-connected test electrodes (lower trace), which can be attributed to residual surface conductivity and is of the same order of the sensitivity of the detection system.

In all the 7 channels, the I-V curves are symmetrical with respect to voltage polarity. Due to the peaked damage profile shown in Figs. 1 and 3, the conductive region can be reasonably supposed to be localized at the end of the ion range, leading to conductive 150 μm long buried channels at an average depth of 3 μm emerging to the surface at their endpoints. As shown in Fig. 4b, the conductance of the channels increases as a function of the ion fluence, indicating an increasing degree of $sp_2$-like amorphisation of the diamond matrix.

In order to elucidate on the transport mechanism, current-voltage characteristics of channel relevant to a ion fluence of $F = 5.2 \cdot 10^{17}$ ions $cm^{-2}$ were measured at temperatures ranging from 80 to 690 K, in steps of 10 K.

The conductance G (defined as the current/voltage ratio) in log-log scale is shown in Fig. 5 at six different measurement temperatures. At low bias voltage, the characteristics are ohmic (the conductances are constant), whereas, for bias voltages higher than 3 V (i.e. average electric fields higher than 200 V/cm) the current behavior is super-linear.

As reported by other authors [19, 3] for damaged diamond, the variable range hopping in localized states near the Fermi level can be considered a possible mechanism for electronic conduction [20]. A general expression for the resulting three-dimensional hopping conductance at low bias voltages can be written as:

(1) $$G(T) = G_0(T) \cdot \exp\left[-\left(\frac{T_0}{T}\right)^{1/4}\right] = G_{00} \cdot T^{-1/2} \cdot \exp\left[-\left(\frac{T_0}{T}\right)^{1/4}\right]$$

where $G_{00}$ represents the temperature-independent prefactor and the weak temperature dependence $T^{-1/2}$ of the Mott prefactor $G_0(T)$ has been made explicit [21].

Moreover, $T_0$ is defined as:

(2) $$T_0 = \left(\frac{512}{9\pi}\right)\frac{\gamma^3}{k_B N(E_F)}$$

where $k_B$ is the Boltzmann constant, $N(E_F)$ is the density of states at the Fermi level and γ is the decay parameter of the localized wave function ($\gamma^{-1}$ is also known as the localization length).

Figure 6 shows the behavior of $G(T) \cdot T^{1/2}$ in the ohmic regime og G as function of $T^{-1/4}$ according to 3D variable-range hopping, following equation (1) in the 80-690 K temperature range.

In the 270-690 K temperature range, the behavior is evidently linear (correlation coefficient from a linear least squares fit to the data of 0.999); the slope of the linear fit leads to a value of $T_0$ of $(2.18 \pm 0.06) \cdot 10^8$ K, which corresponds to the localization parameter $N(E_F) \cdot \alpha^{-3} = (9.5 \pm 0.3) \cdot 10^{-4}$ $eV^{-1}$. Assuming a localization length $\alpha^{-1}$ of 1.2 nm [3, 12], the density of states at the Fermi level is $N(E_F) = 5.5 \cdot 10^{17}$ $cm^{-3} \cdot eV^{-1}$.

This value can be compared with the early study of Hauser on the effects of 20-70 keV C ion damage in diamond [3]. From such report, it is possible to evaluate the C implantation fluence that results in the same value of $N(E_F)$ reported here, which is $F \cong 5 \cdot 10^{15}$ ions $cm^{-2}$, two order of magnitude smaller of He fluence reported in our work for the channel under investigation (F = $5.2 \cdot 10^{17}$ ions $cm^{-2}$). Such a significant discrepancy can be fully reconciled if two relevant aspects are empirically taken into account. Firstly, the maximum damage density of 70 keV C ions is ~10 times higher than for 1.8 MeV He ions, as shown in Fig. 1, thus accounting for a factor ~10 in the fluences ratio. Moreover, the ion damage process in diamond is known to be more effective for shallow implantations in comparison to deep implantations. This fact is to be attributed to the higher internal pressure associated with the significant volume expansion of the damaged layer in the case deep implantations, which results in a higher inertness to structural phase transformations [22, 23]. This difference has been quantitatively estimated by comparing the values of the graphitization thresholds for shallow and deep implantations. In previous works, values of $1 \cdot 10^{22}$ vacancies $cm^{-3}$ and $9 \cdot 10^{22}$ vacancies $cm^{-3}$ were reported for shallow and deep implantations, respectively [4, 24]. Therefore, the remaining difference of one order of magnitude in the above mentioned damage densities can be qualitatively attributed to the ~9 ratio in the respective graphitization thresholds, which qualitatively account for the resilience of the diamond lattice to structural damage at different depths.

At temperatures lower than 250 K, the curve in figure 6 shows a remarkable deviation from the linear behavior as follows from eq. (1).

To examine the field enhanced transport mechanism occurring at bias voltages higher than 3 V (see fig. 5), it is more appropriate to consider the current-voltage characteristics shown in Fig. 7. In the log-log scale the behaviors are clearly linear (correlation coefficients higher than 0.999) at all the temperatures in the 250-690 K range.

The power dependence of the current on the applied bias voltage is consistent with a space charge limited current behavior occurring at the low-moderate electric fields ranging from 200 to 3000 $V \cdot cm^{-1}$. The slope of the log(I)-log(V) curves, i.e. the exponent α of the power law $I \propto V^{\alpha}$ , is influenced by the type of trap distribution in the bandgap [25] and presents a weak dependence from the temperature as shown in Fig. 8. As the temperature increases, the exponent α decreases, suggesting a progression to an increasingly ohmic behavior (i.e. α=1). This trend is consistent with the results of our previous work [6], which evidenced the phase transition to graphite and a complete ohmic behavior after annealing at 800°C of buried channels, with an induced damage density higher than the critical fluence threshold.

## Conclusions

Buried conductive channels were realized in single diamond using 1.8 MeV He focused ion beams to induce a transformation of diamond to a conducting form of carbon. A novel and simple method to make electrical contacts to the buried layers was employed: W wires act as variable thickness masks to reduce the depth of the damaged regions, allowing the surface contacting of the channel terminations without any further fabrication stage [26].

The conductance at room temperature of the buried channels monotonically increases as function of the implanted fluence in the range $2.1 \cdot 10^{16}$ to $6.3 \cdot 10^{17}$ ions/cm$^2$ In the ohmic regime the conductance vs. temperature data, in the range of 270-690 K, are consistent with the variable range hopping transport mechanism.

The density of states at the Fermi level ($N(E_F) = 5.5 \cdot 10^{17}$ cm$^{-3} \cdot$eV$^{-1}$ at a fluence of $5.2 \cdot 10^{17}$ ions$\cdot$cm$^{-2}$) is comparable with those obtained in previous works, with heavier ions at lower energy, if both the relevant damage profile and the dependence of the graphitization threshold on the implantation depth are properly taken into account.

For electric fields higher than 200 V/cm, at any temperature in the 250-690 K range, the current shows a superlinear behaviour on the applied bias voltage. Although numerous models [27] can be adopted to fit the experimental data in the limited electric field range reported in this work, the well-defined power law dependence of the current from the applied bias voltage suggest a space charge limited current transport influenced by the trap distribution and type, as highlighted by the weak dependence of the power exponent on the temperature.

## Acknowledgments

The work of P. Olivero is supported by the “Accademia Nazionale dei Lincei – Compagnia di San Paolo” Nanotechnology grant, which is gratefully acknowledged.

**Figures and captions**

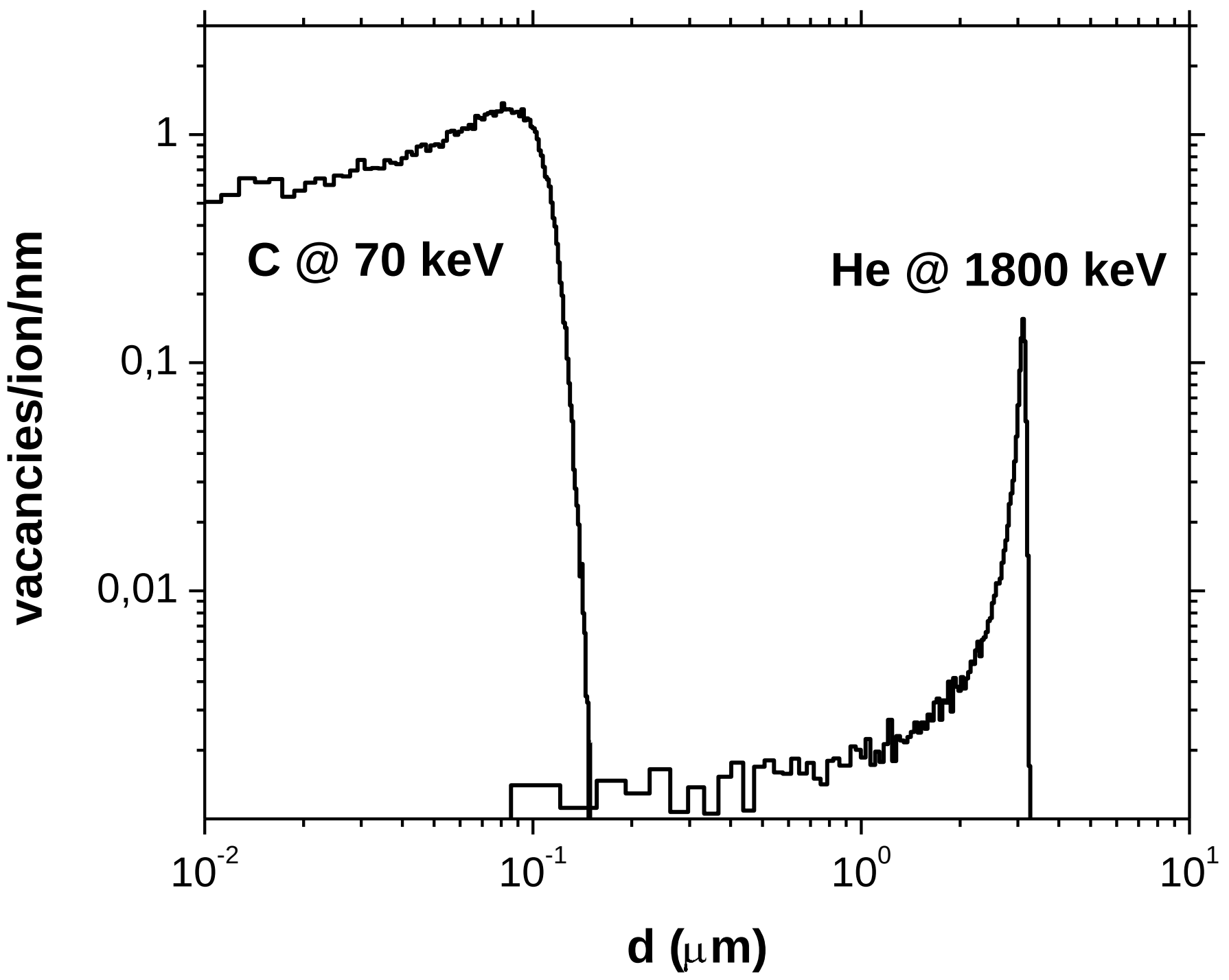

Fig. 1: SRIM Monte Carlo simulations of the damage density profile induced in diamond by 70 keV C ions [3] and 1.8 MeV $He^+$ ions (present work); the curves were calculated by setting a value of 50 eV for the atom displacement energy in the diamond lattice [18].

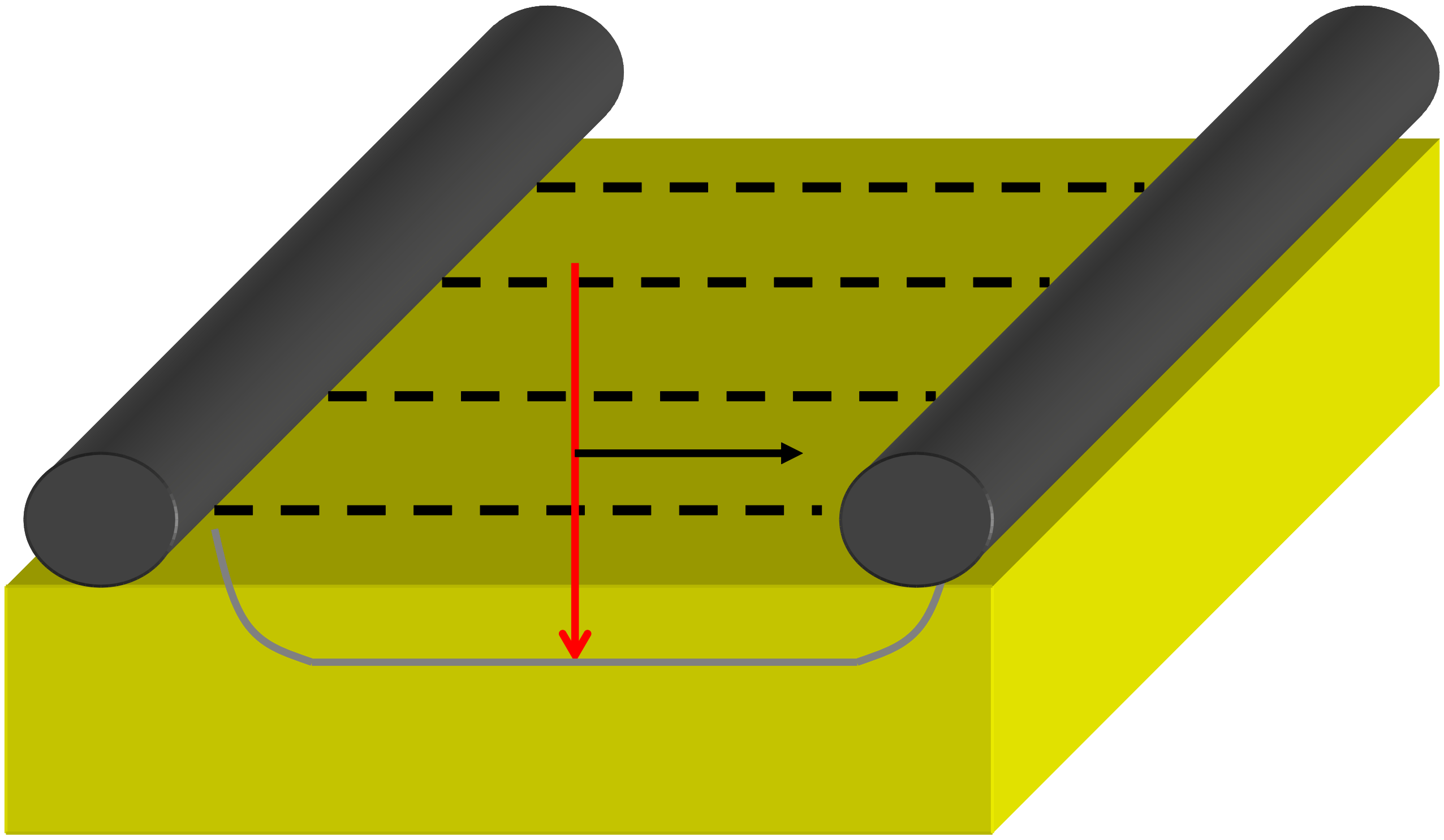

Fig. 2 (color online): Schematics of the three-dimensional masking technique adopted to control the penetration depth of implanted ions. The ion microbeam scanning between the edges of the wires (in grey) perpendicularly to the sample surface is represented as the vertical red arrow. The damaged layer at a variable depth in the material is represented by the grey line.

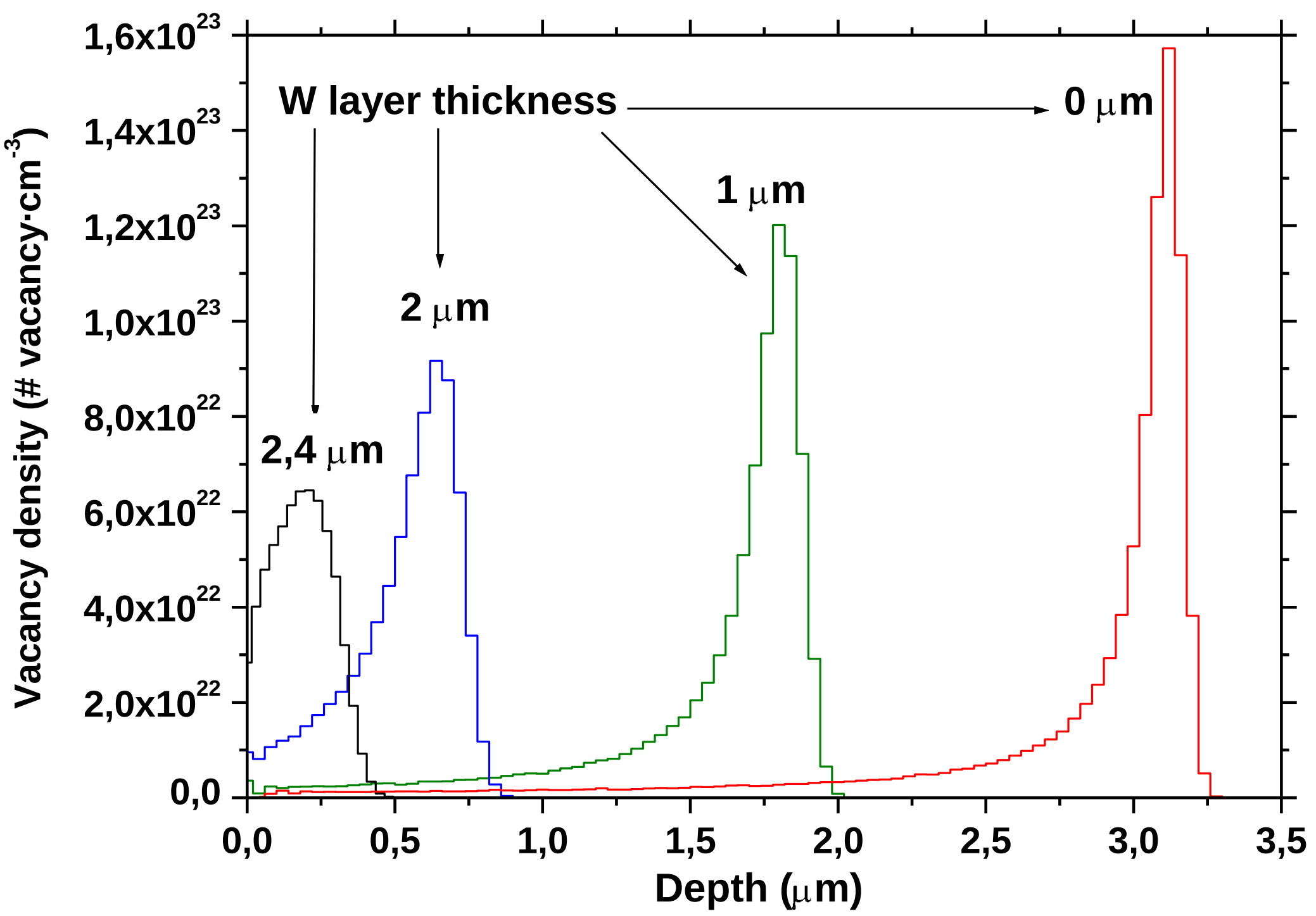

Fig. 3 (color online): SRIM evaluation of vacancy profiles in diamond by 1.8 MeV $He^+$ ions after crossing different tungsten thicknesses. The vacancy concentration has been evaluated by multiplying the Vacancy/ion profile (see Fig. 1) by the ion fluence of $10^{17}$ ions/cm$^2$.

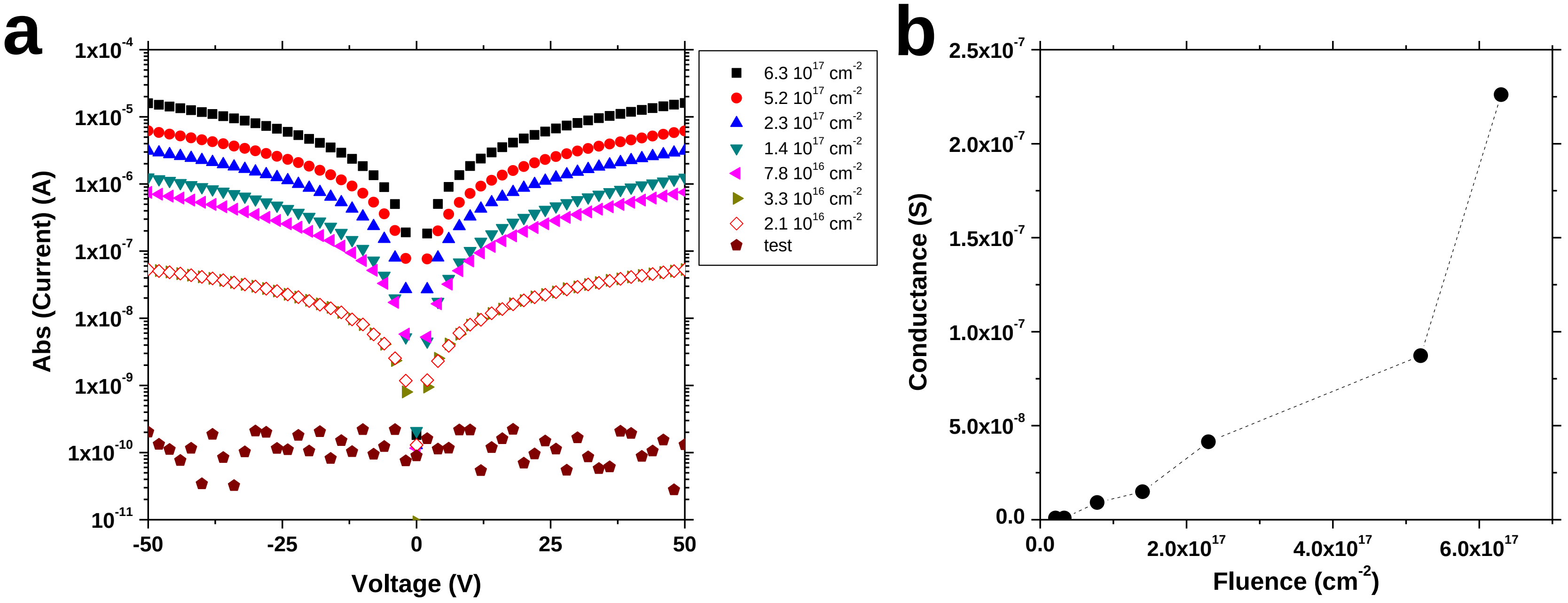

Fig. 4 (color online): a) current-voltage characteristics of the 7 regions damaged at increasing fluence; the last dataset (labelled as "test") refers to the IV characteristic between two electrodes not connected by damaged channels; b) conductance of the channels evaluated at V = 2 V as a function of the implantation fluence.

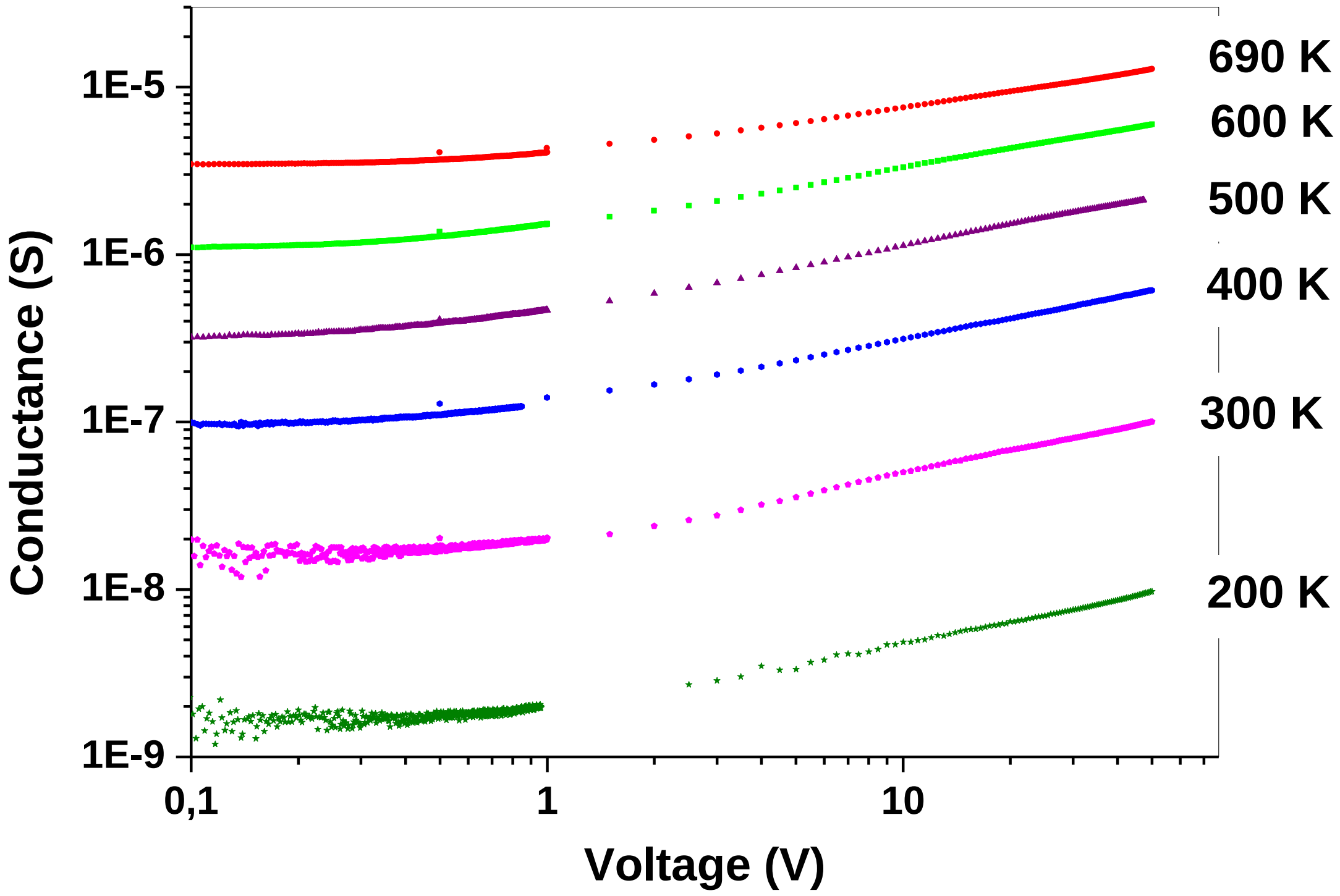

Fig. 5: current-voltage characteristics of the channel implanted at a fluence of $5.2 \cdot 10^{17}$ $cm^{-2}$ at 6 different measurement temperatures.

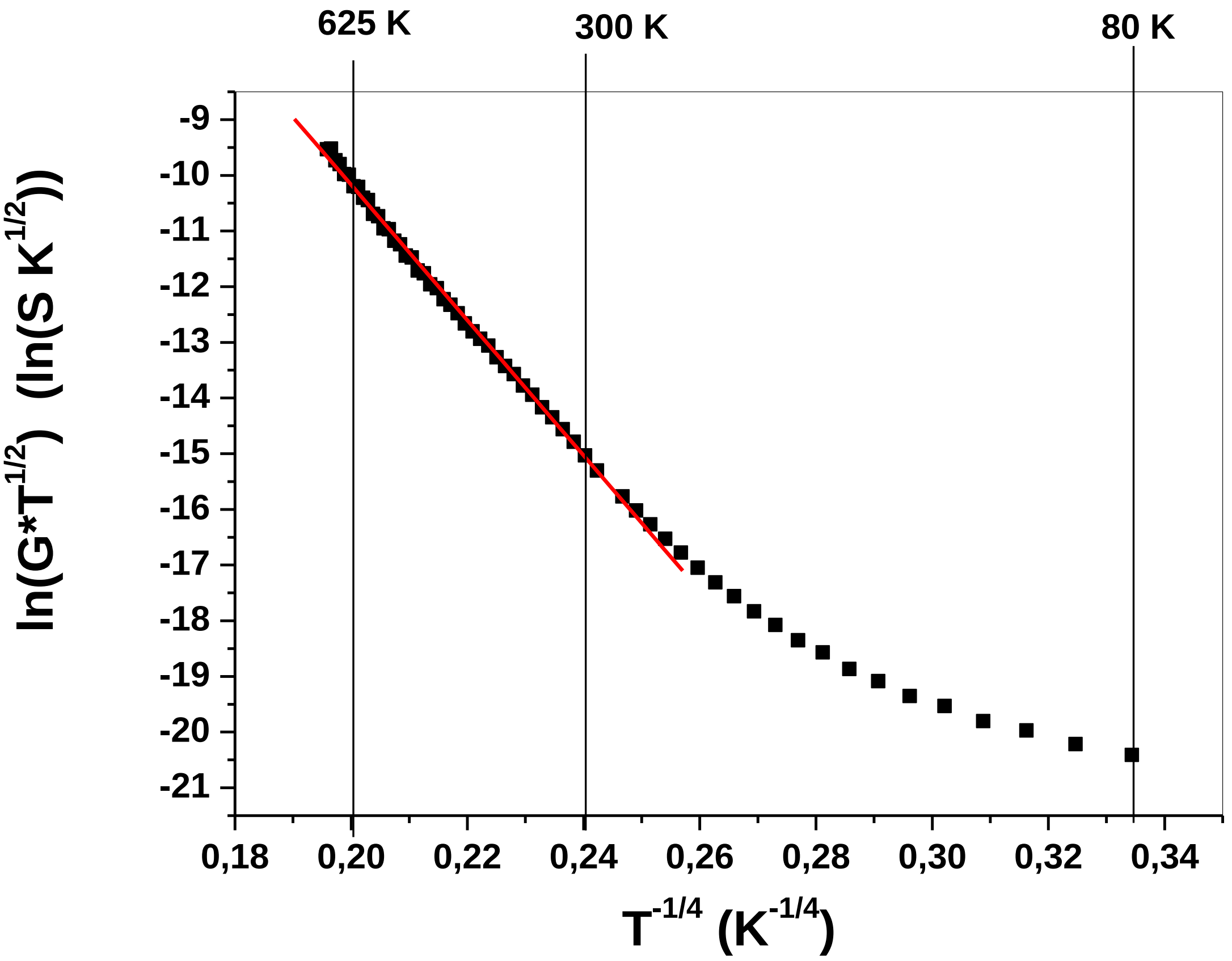

Fig.6: Behavior of the conductance G in the ohmic regime vs $T^{-1/4}$ (Eq. 1). The line is the linear fit in the temperature range 270-690 K.

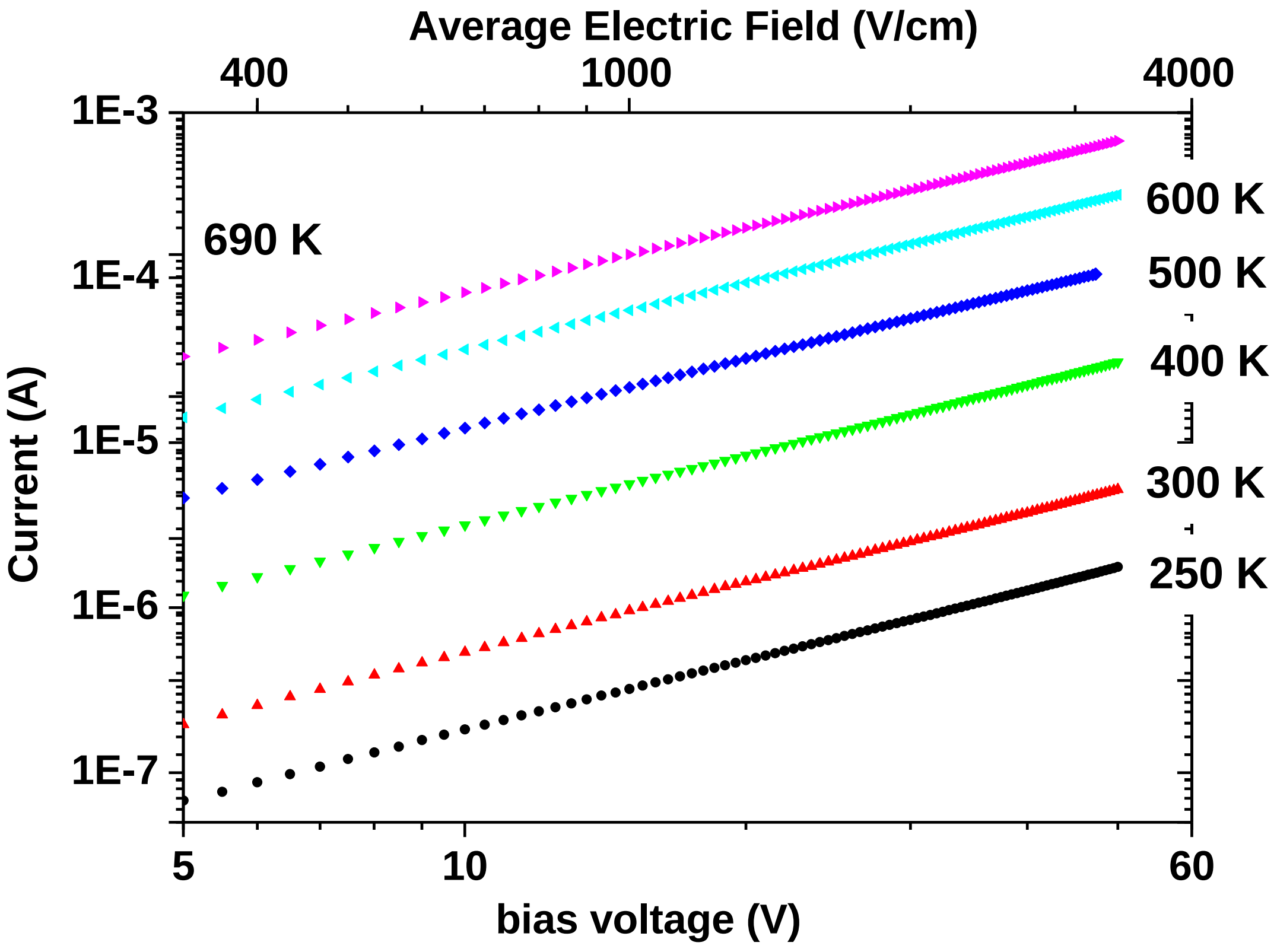

Fig. 7: log-log plot of the current-voltage characteristics at different temperatures.

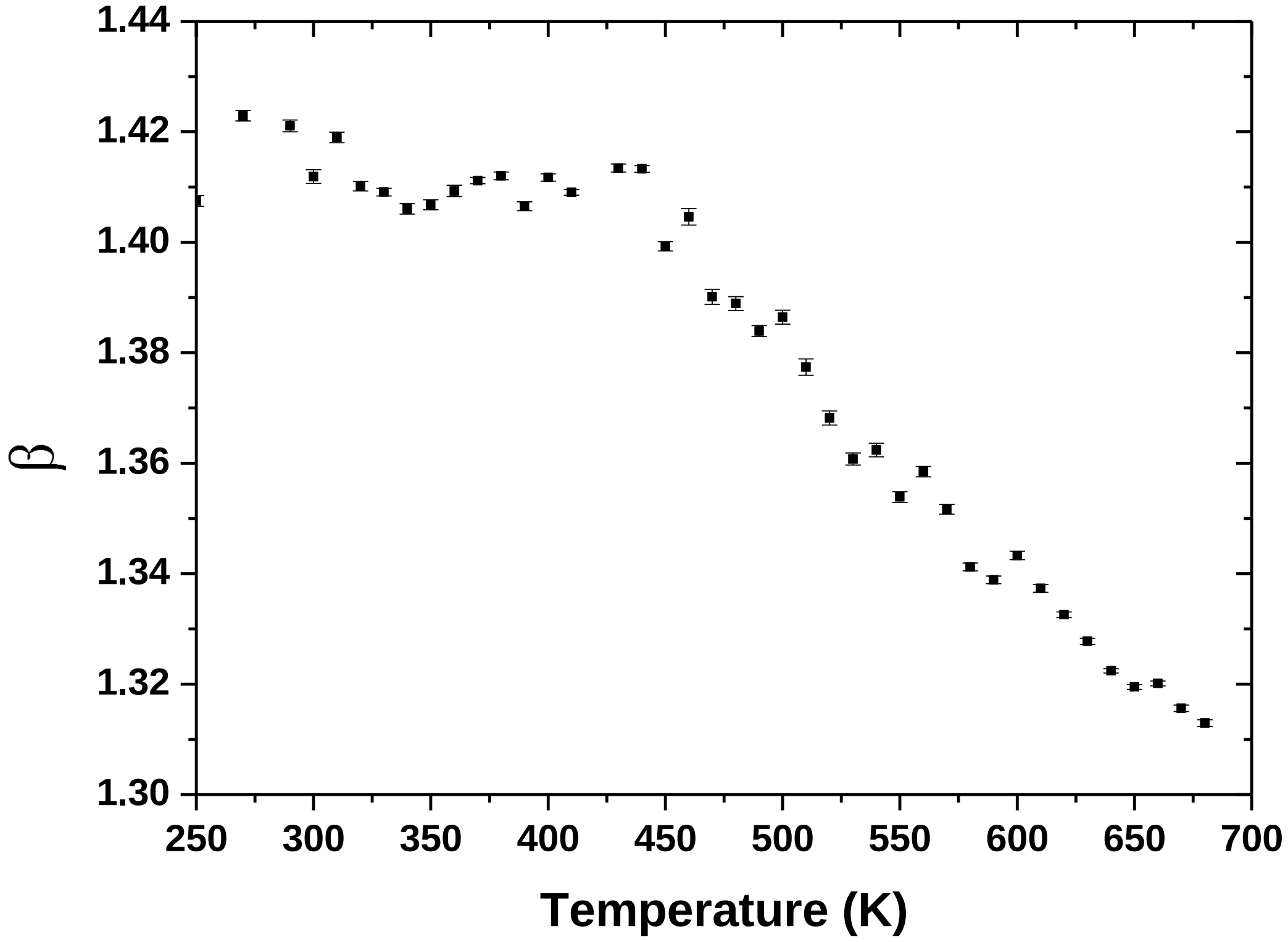

Fig. 8: dependence of the normalized differential conductance $\alpha = \frac{d\ln(I)}{d\ln V}$ [25] on the temperature. The error bars are relevant to the linear fitting procedure of the data shown in Fig. 7.